# Dissipation losses limiting first-order phase transition materials in cryogenic caloric cooling: A case study on all-d-metal Ni(-Co)-Mn-Ti Heusler alloys

Benedikt Beckmann[a,*], David Koch[b], Lukas Pfeuffer[a], Tino Gottschall[c], Andreas Taubel[a], Esmaeil Adabifiroozjaei[d], Olga N. Miroshkina[e], Stefan Riegg[a], Timo Niehoff[c,f], Nagaarjhuna A. Kani[a,d], Markus E. Gruner[e], Leopoldo Molina-Luna[d], Konstantin P. Skokov[a] and Oliver Gutfleisch[a]

[a]*Functional Materials, Institute of Materials Science, Technical University of Darmstadt, Darmstadt 64287, Germany*
[b]*Structure Research, Institute of Materials Science, Technical University of Darmstadt, Darmstadt 64287, Germany*
[c]*Dresden High Magnetic Field Laboratory (HLD-EMFL), Helmholtz-Zentrum Dresden-Rossendorf (HZDR), Dresden 01328, Germany*
[d]*Advanced Electron Microscopy (AEM), Institute of Materials Science, Technical University of Darmstadt, Darmstadt 64287, Germany*
[e]*Faculty of Physics and Center for Nanointegration Duisburg-Essen (CENIDE), University of Duisburg-Essen, Duisburg 47057, Germany*
[f]*Institute of Solid State and Materials Physics, Technische Universität Dresden, Dresden 01069, Germany*

ABSTRACT

Ni-Mn-based Heusler alloys, in particular all-d-metal Ni(-Co)-Mn-Ti, are highly promising materials for energy-efficient solid-state refrigeration as large multicaloric effects can be achieved across their magnetostructural martensitic transformation. However, no comprehensive study on the crucially important transition entropy change $\Delta s_t$ exists so far for Ni(-Co)-Mn-Ti. Here, we present a systematic study analyzing the composition and temperature dependence of $\Delta s_t$. Our results reveal a substantial structural entropy change contribution of approximately 65 J(kgK)$^{-1}$, which is compensated at lower temperatures by an increasingly negative entropy change associated with the magnetic subsystem. This leads to compensation temperatures $T_{comp}$ of 75 K and 300 K in Ni$_{35}$Co$_{15}$Mn$_{50-y}$Ti$_y$ and Ni$_{33}$Co$_{17}$Mn$_{50-y}$Ti$_y$, respectively, below which the martensitic transformations are arrested. In addition, we simultaneously measured the responses of the magnetic, structural and electronic subsystems to the temperature- and field-induced martensitic transformation near $T_{comp}$, showing an abnormal increase of hysteresis and consequently dissipation energy at cryogenic temperatures. Simultaneous measurements of magnetization and adiabatic temperature change $\Delta T_{ad}$ in pulsed magnetic fields reveal a change in sign of $\Delta T_{ad}$ and a substantial positive and irreversible $\Delta T_{ad}$ up to 15 K at 15 K as a consequence of increased dissipation losses and decreased heat capacity. Most importantly, this phenomenon is universal, it applies to any first-order material with non-negligible hysteresis and any stimulus, effectively limiting the utilization of their caloric effects for gas liquefaction at cryogenic temperatures.

## 1. Introduction

In an age when anthropogenic climate change and depletion of natural energy resources are two of the most urgent world-wide societal and existential challenges, the search for energy-efficient and environmentally friendly technologies is of outermost importance. This development is strongly intertwined with globally rising population and standard of living [1], leading to an ever growing demand for cooling [2, 3], which is predicted to exceed the energy consumption for heating in this century [4]. Currently, most room-temperature cooling devices are based on the 120-year old vapor-compression refrigeration cycle, which lacks in thermodynamic efficiency and has a profound negative environmental impact [5]. In addition, the compression-based cooling process used for the liquefaction of hydrogen, which is critical as an alternative energy carrier for the transition towards renewable energies [6, 7], is considered to be inefficient as well [8, 9]. Therefore, future cooling technologies for room as well as cryogenic temperature applications need to provide energy-efficient and environmentally friendly alternatives.

As such an alternative, solid-state caloric cooling is based on the response of caloric materials to the application of external stimuli, such as electric or magnetic fields [10], hydrostatic pressure [11, 12] and uniaxial stress [13]. Multicaloric materials are susceptible to more than one stimulus [14, 15], allowing the design of advanced cooling cycles, e.g. by subsequently utilizing magnetic field and uniaxial stress [16]. Depending on the selected thermodynamic boundary conditions, the stimulus leads to an isothermal entropy change $\Delta s_T$ or adiabatic temperature change $\Delta T_{ad}$ of the material. Conventional caloric materials respond with $\Delta s_T < 0$ and $\Delta T_{ad} > 0$ to the application of a given external stimulus, whereas inverse caloric materials show $\Delta s_T > 0$ and $\Delta T_{ad} < 0$ [14]. Of all ferroic cooling technologies, magnetocalorics is considered to be the best studied [17, 18] and shows an improved thermodynamic efficiency compared to the widely used compression-based technology [19, 20].

*Corresponding author
✉ benedikt.beckmann@tu-darmstadt.de ( Beckmann)
ORCID(s): 0000-0002-2232-1804 ( Beckmann)

However, the search for promising magnetocaloric materials is still ongoing.

Since the discovery of the giant magnetocaloric effect in $Gd_5Si_2Ge_2$ [21], various material systems showing enhanced magnetocaloric properties due to first-order phase transitions have been discovered [22]. The most prominent material systems are $La(Fe,Si)_{13}$ [23, 24], $Fe_2P$-type compounds [25–27], Ni-Mn-based Heusler alloys [10, 28–31] and in some extend Fe-Rh [32–34]. In Ni-Mn-based Heusler alloys, giant caloric effects are coupled to first-order magnetostructural martensitic transformations between low-temperature martensite and high-temperature austenite. In this work, multicaloric all-d-metal Ni(-Co)-Mn-Ti Heusler alloys are investigated as these alloys have recently attracted particular interest in the fields of magneto- [35–38], elasto- [39–42], baro- [43, 44] and multicalorics [45] due to an improved mechanical stability [39, 40], large volume changes [46] as well as good tuneability of the martensitic transformation temperature $T_t$ and austenite Curie temperature $T_C^A$ [38, 46, 47].

In general, knowledge about composition and temperature dependencies of the transition entropy change $\Delta s_t$ in any given caloric material is crucial to develop samples with tailored properties for solid-state cooling applications. Thereby, independent on the selected external stimulus to drive the phase transition, the transition entropy change is of universal importance and is expressed as

$$\Delta s_t = \Delta s_{lat} + \Delta s_{mag} + \Delta s_{el} \qquad (1)$$

with $\Delta s_{lat}$, $\Delta s_{mag}$ and $\Delta s_{el}$ being the entropy change contributions associated with the structural, magnetic and electronic subsystem, respectively. Kihara *et al.* showed that the entropy change of martensitic transformations in Heusler alloys is dominated by the structural subsystem, whereas the contribution of the electronic subsystem is negligibly small [48], i.e. $|\Delta s_{lat}| > |\Delta s_{mag}| \gg |\Delta s_{el}|$. In inverse magnetocaloric Heusler alloys, such as Ni(-Co)-Mn-Ti, the dominating $\Delta s_{lat}$ is positive and $\Delta s_{mag}$ is increasingly negative towards lower temperatures. The opposing positive lattice and negative magnetic entropy change contributions give rise to the "dilemma of inverse magnetocaloric materials" [49] and are compensated at the compensation temperature $T_{comp}$, below which the martensitic transformation is thermally arrested due to the absence of a driving force. In the literature, this thermodynamic effect is often called kinetic arrest and is described in multiple classical Ni(-Co)-Mn-X Heusler alloys with X being a main group element, namely Sn [50], Sb [51], Al [52] and In [49, 53–55]. However, the arrest phenomenon is not only limited to Heusler alloys but also appears in other inverse magnetocaloric materials, such as Fe-Rh compounds [34].

In this study, we disentangle the contributions to the transition entropy change of the magnetostructural martensitic transformation in multicaloric all-d-metal Ni(-Co)-Mn-Ti Heusler alloys, since there is so far no comprehensive study on $\Delta s_t$ in this material system, even though the transition entropy change is a fundamental material property governing all caloric effects. On this basis, we analyze the responses of the magnetic, structural and electronic subsystems to the temperature- and field-induced martensitic transformation, showing an abnormally increased magnetic hysteresis width at temperatures near and below $T_{comp}$. Based on this, we reveal the detrimental effect of non-negligible hysteresis and associated dissipation losses of first-order phase transitions on the adiabatic temperature change, representing a crucial limitation for recently emerging research interest in caloric cooling applications at cryogenic temperatures, such as hydrogen liquefaction [56–58].

## 2. Experimental Details
### 2.1. Synthesis

Various $Ni_{50-x}Co_xMn_{50-y}Ti_y$ (x=0,15,17 & 6≤y≤17) samples have been synthesized by arc melting high-purity elements in protective Ar atmosphere. Due to evaporation losses, 3% excess Mn was added. Each 10 g sample was turned and remolten at least five times to ensure a homogeneous distribution of elements. Based on the optimized heat treatment conditions found in reference [38], the ingots have been annealed at 1323 K for 96 h in sealed quartz ampules with Ar atmosphere. Subsequently, the samples were quenched by breaking the quartz tube in water.

### 2.2. Microstructural and structural characterization

The crystal structure and phase-purity have been characterized by powder X-Ray diffraction (XRD) in transmission geometry. Room-temperature XRD has been carried out with a *Stoe* Stadi P diffractometer using Mo $K_{\alpha 1}$ radiation and a position sensitive detector in a $2\theta$ range from 5° to 50° with an effective stepsize of 0.01°. Temperature-dependent XRD has been performed with a purpose-built diffractometer with Mo $K_{\alpha 1}$ radiation and a *Dectris* Mythen 1K R silicon strip detector in a $2\theta$ range from 7° to 58° with an effective step size of 0.009°. A detailed description of the device can be found in reference [59]. In this setup, powder samples have been mixed with *NIST* SRM 640d silicon standard powder and have been glued on graphite foil. The annealed arc molten bulk samples have been milled to powder of particle size <80 μm. To ensure the release of milling-induced stresses, the powder was recrystallized by annealing. Structural analysis has been performed by Rietveld-refinement using *FullProf* software package [60] for austenite and unmodulated martensite and *JANA2006* [61] for modulated martensite. Images of the crystal structures have been created with *VESTA* [62].

The microstructure and chemical composition have been analyzed with backscatter electron (BSE) imaging and energy-dispersive X-Ray spectroscopy (EDX) using a *Tescan* Vega3 scanning electron microscope (SEM). A *Zeiss* Axio Imager.D2M has been used to obtain optical microscopy micrographs of austenite and martensite. The nanostructure of selected samples has been characterized at room-temperature based on high-resolution transmission electron microscopy (HRTEM) images and selected area electron diffraction (SAED) patterns obtained with the 200 kV *JEOL* JEM 2100-F transmission electron microscope (TEM). The TEM lamellas have been prepared by a two-step ion milling process at 80 K using a *Gatan* 691 Precision Ion Polishing System (PIPS) following the procedure described in reference [63].

### 2.3. Magnetometry

Magnetic measurements have been performed with a *LakeShore* vibrating sample magnetometer (VSM) and *Quantum Design* physical property measurement system (PPMS-14T). Isofield measurements have been performed with a heating and cooling rate of 2 Kmin$^{-1}$. Isothermal measurements have been carried out with a magnetic field ramp rate of 5 mTs$^{-1}$. To erase the remnants of the magnetic field-induced phase transformation, a discontinuous temperature protocol was used after each isothermal magnetic measurement. In such a protocol, the sample is heated above and subsequently cooled below the transition temperature in zero-field. Based on the field-dependent magnetic measurements, the isothermal entropy change $\Delta s_T(T, H)$ of the field-induced martensitic transformation has been calculated following the guidelines in references [64, 65] using the Maxwell-relation

$$\Delta s_T(T, H) = \mu_0 \int_0^H \left(\frac{\partial M}{\partial T}\right)_H dH. \quad (2)$$

Based on isofield magnetic measurements, the transition entropy change $\Delta s_t$ has been estimated with Clausius-Clapeyron (CC) equation

$$\frac{\Delta M}{\Delta s_t} = \frac{dT_t}{\mu_0 dH} \quad (3)$$

using the magnetization change $\Delta M$ across the phase transformation in 1 T as well as the sensitivity of the phase transformation towards the magnetic field stimulus $dT_t/\mu_0 dH$, determined based on $M(T)$ measurements in 0.1, 1, and 2 T. The transition entropy changes estimated with the Clausius-Clapeyron equation are labeled as $\Delta s_{t,CC}$ in the following.

To monitor the responses of all three material subsystems to the martensitic transformation close to and below the compensation temperature, simultaneous measurements of magnetization $M$, strain $\Delta L/L_0$ and electrical resistivity $\rho$ have been carried out using a purpose-built insert for the PPMS-14T VSM option. To minimze the influence of texture and to increase statistical robustness, the strain has been determined as the average strain detected by two strain gauges glued to the sample surface parallel and perpendicular to the magnetic field direction. The temperature- and field-induced response of the strain gauges itself has been corrected by simultaneously measuring two reference strain gauges glued to the quartz sample holder. The electrical resistivity has been determined with the two-contact method. A detailed description of the device can be found in reference [66].

### 2.4. Calorimetry

Magnetic field-dependent heat capacity $c_p(T, H)$ has been measured with the PPMS-14T in magnetic fields of 0, 2, 5, and 14 T in a temperature range from 2 K to 395 K. The total entropy $s(T, H)$ of the material has been calculated as

$$s(T, H) = \int_{2K}^{T} \left(\frac{c_p}{T}\right)_H dT. \quad (4)$$

Magnetic field-dependent transition entropy changes calculated based on this measurement are labeled $\Delta s_{t,c_p}^{H_i}$ in the following.

Differential scanning calorimetry (DSC) measurements have been performed with a *Netzsch* DSC 404 F1 Pegasus in a temperature range from 150 K to 700 K and a heating and cooling rate of 5 Kmin$^{-1}$. For this purpose, Al crucibles have been used in the silver-furnace setup, equipped with a liquid nitrogen cooling system. The transition entropy change $\Delta s_t$ of the reverse martensitic transformation, i.e. the martensite to austenite transformation, has been determined by integrating the baseline-corrected mass-specific heat flow $\dot{Q}$ over the phase transformation region

$$\Delta s_t = \int_{A_s}^{A_f} \left(\dot{Q} - \dot{Q}_{Baseline}\right) T^{-1} \left(\frac{dT}{dt}\right)^{-1} dT \quad (5)$$

with $A_s$ and $A_f$ being the start and finish temperatures of the transformation. This procedure was verified for selected samples by calculating the zero-field heat capacity based on additional calibration measurements using a sapphire as heat capacity standard. To distinguish both methods, $\Delta s_{t,\dot{Q}}^{0T}$ and $\Delta s_{t,c_p}^{0T}$ are used in the following.

### 2.5. Adiabatic temperature change measurements

Adiabatic temperature changes $\Delta T_{ad}$ associated with the magnetocaloric effect have been simultaneously measured with the magnetization of the sample in a solenoid magnet at the Dresden High Magnetic Field Laboratory (HLD) in pulsed magnetic fields of 10, 14, 20, 30, and 50 T. The magnetic field was determined by means of a calibrated pick-up coil. The maximum field strength of each pulse was always reached after 13 ms. The adiabatic temperature change has been measured with a thin type T thermocouple glued between two pieces of the sample with silver epoxy [67]. The magnetization of the sample has been measured simultaneously with a compensated split coil wound around the sample and a non-magnetic counterpart with opposite winding (see supplementary material S1). The voltage signal was fine compensated numerically by a small correction using the field-induced signal of the pick coil and then integrated to dimensionless magnetization.

Adiabatic temperature changes associated with the elastocaloric effect have been detected in an *Instron* 5967 30 kN universal testing machine, equipped with a temperature chamber. Strain and force have been monitored with a strain gauge extensometer attached to the compression platens close to the specimen and load cell, respectively. To ensure quasi-adiabatic conditions, strain rates of $3 \times 10^{-2}$ s$^{-1}$ have been used [63]. The $\Delta T_{ad}$ has been measured with a K-type thermocouple attached to the surface of the sample.

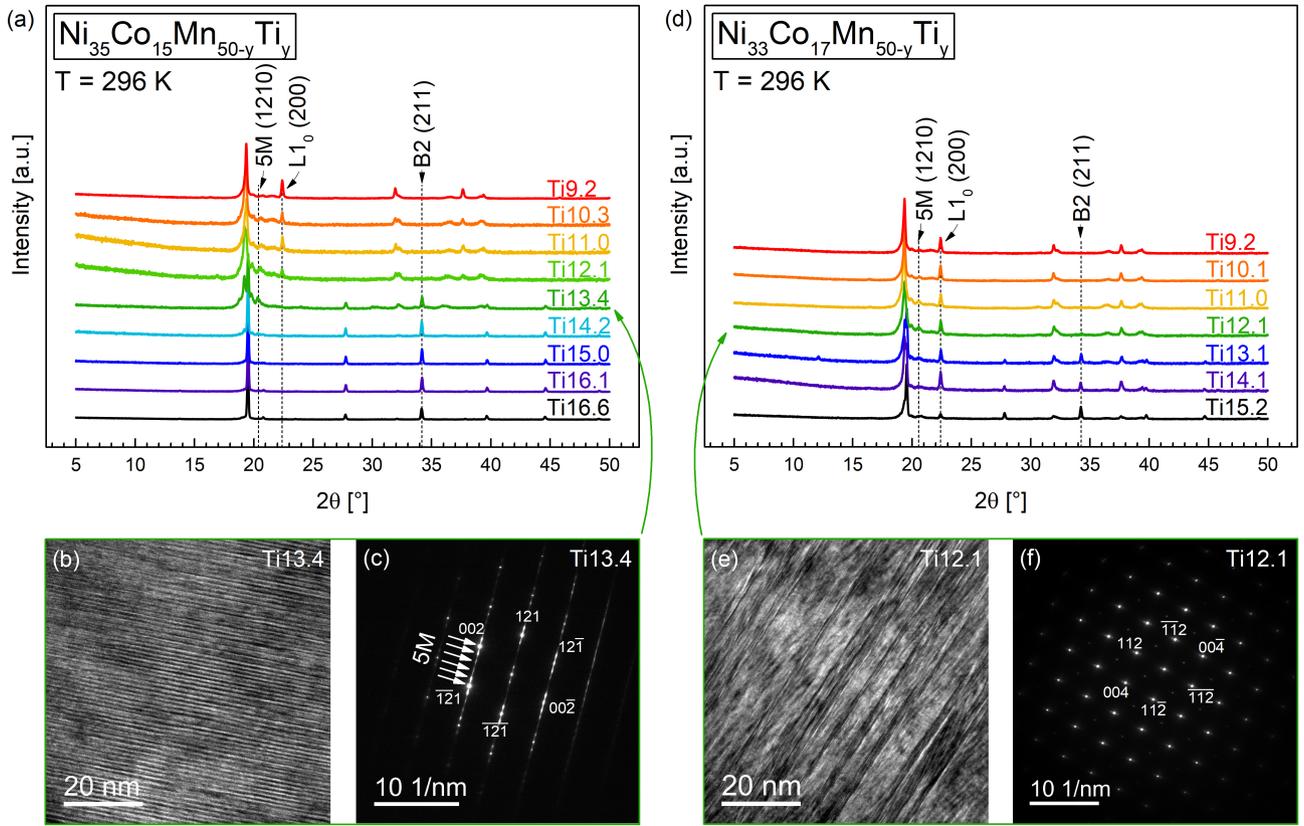

**Figure 1:** Room-temperature XRD patterns of annealed nominal $Ni_{35}Co_{15}Mn_{50-y}Ti_y$ (a) and $Ni_{33}Co_{17}Mn_{50-y}Ti_y$ (d) powders showing the gradual transition from cubic B2 austenite to a mixture of incommensurately modulated monoclinic and non-modulated tetragonal $L1_0$ to single phase $L1_0$ martensite crystal structures (see vertical dashed lines) with decreasing Ti content. The intensity of each pattern has been normalized. The patterns are labeled according to the Ti content determined by EDX. HRTEM images (b, e) and SAED patterns (c, f) show 5M modulated and non-modulated $L1_0$ martensite for $Ni_{33.9}Co_{14.6}Mn_{38.1}Ti_{13.4}$ (b, c) and $Ni_{31.6}Co_{16.6}Mn_{39.7}Ti_{12.1}$ (e, f), respectively. The SAED patterns of the 5M modulated and $L1_0$ non-modulated martensite are obtained along the $[\bar{4}20]$ and $[1\bar{1}0]$ zone axis, respectively.

### 2.6. Density functional theory (DFT)

To take a closer look at the effect of Co on the magnetic properties, DFT calculations have been performed for the ferromagnetic cubic B2 austenite with lattice parameters taken from XRD experiments. Chemical disorder was modeled analytically within the coherent potential approximation (CPA) allowing to simulate disordered structures in small unit cells. The total magnetization and exchange constants were obtained with the help of Korringa–Kohn–Rostoker (KKR) approach as implemented in the Munich SPR-KKR code [68, 69] in full-potential mode together with scalar relativistic corrections. The exchange-correlation functional was treated within the generalized gradient approximation (GGA) following the Perdew, Burke, and Ernzerhof (PBE) scheme [70]. The angular momentum expansion was carried out up to $l_{max} = 3$ ($f$-states). We assumed electronic self-consistency to be reached when the error in the potential functions dropped below $10^{-6}$. Brillouin zone integration was carried out using the special point method with a regular $k$-point grid of 1540 points, which corresponds to a 39×39×39 mesh in the full Brillouin zone. The Heisenberg model exchange parameters $J_{ij}$ between pairs of atoms $i$ and $j$ of all different chemical types and positions within a cluster radius of four lattice constants were calculated following Liechtenstein's approach [71].

## 3. Results and Discussion

### 3.1. Structural analysis

Room-temperature XRD patterns of nominal and chemically homogeneous all-d-metal $Ni_{35}Co_{15}Mn_{50-y}Ti_y$ (9≤y≤17) and $Ni_{33}Co_{17}Mn_{50-y}Ti_y$ (9≤y≤15) Heusler alloys are shown in figure 1 (a) and (d), respectively. In good agreement with [47], a gradual transition from cubic B2 austenite to a mixture of incommensurately modulated monoclinic and non-modulated tetragonal $L1_0$ to single phase $L1_0$ martensite is observable with decreasing Ti content in both series. This indicates the presence and tuneability of martensitic transformations by composition in both series. No additional phases besides martensite and austenite can be detected in all samples. The increasing volume fraction of $L1_0$ martensite with decreasing Ti content is identical to classical Heusler alloys, such as Ni-Mn-Ga [72], Ni-Mn-Sn [73, 74] and Ni-Mn-In [75]. The B2 ordered austenite is determined by the absence and presence of the (111) and (200) reflections, respectively. The presence of B2 order is in good agreement with recent results of neutron diffraction experiments [76] and is seen as the origin of the dependence of the austenite Curie temperature $T_C^A$ on the Ti content in this material system [38].

Figure 1 (b, e) and (c, f) show HRTEM images and corresponding SAED patterns of one martensitic sample of each Co-series at room-temperature, respectively. The $Ni_{33.9}Co_{14.6}Mn_{38.1}Ti_{13.4}$ sample (see figure 1 (b, c)) exhibits 5M modulated martensite as indicated by the satellite reflexes caused by the modulation. The TEM analysis of $Ni_{31.6}Co_{16.6}Mn_{39.7}Ti_{12.1}$ (see figure 1 (e, f)) reveals non-modulated $L1_0$ martensite. Therefore, both SAED patterns are in good agreement with the powder XRD results.

### 3.2. Transition entropy change

To study the influence of compositional variations on the magnetostructural martensitic transformations in all-d-metal Ni(-Co)-Mn-Ti Heusler alloys in detail, we performed isofield magnetization measurements in 1 T of nominal $Ni_{35}Co_{15}Mn_{50-y}Ti_y$ (9≤y≤17) and $Ni_{33}Co_{17}Mn_{50-y}Ti_y$ (9≤y≤15) samples, shown in figure 2 (a) and (d), respectively. The martensitic transformations are clearly observable due to the magnetization change accompanying the transition between low-temperature weak-magnetic martensite and high-temperature para- or ferromagnetic austenite and are in good agreement with room-temperature XRD results (see figure 1 (a) and (d)). Larger magnetization changes can be achieved at lower temperatures within each Co-series as the austenite saturation magnetization rises with decreasing temperature and $T_C^A - T_t$ grows with increasing Ti content. Since DFT calculations show that the addition of Co enhances saturation magnetization and ferromagnetic coupling (see supplementary material S2), larger $\Delta M$ are present in $Ni_{33}Co_{17}Mn_{50-y}Ti_y$ compared to $Ni_{35}Co_{15}Mn_{50-y}Ti_y$ at any given transition temperature. As it will be shown later, the increased low-temperature magnetization of samples with a low transition temperature is linked to the presence of residual austenite. The composition dependencies of the martensitic transformation temperature $T_t$ and austenite Curie temperature $T_C^A$ can be well described with the average number of valence electrons per atom $e/a$ at $T \geq T_{comp}$, as demonstrated in reference [38]. However, the estimation of $T_t$ based on the $e/a$ ratio breaks down at $T = T_{comp}$ as no temperature-induced martensitic transformations can be observed below approximately 75 K and 300 K in samples with a Ti content larger than 16.1 at.% and 12.1 at.% in the $Ni_{35}Co_{15}Mn_{50-y}Ti_y$ and $Ni_{33}Co_{17}Mn_{50-y}Ti_y$ series, respectively. This is in good agreement with [47] in which no transformation has been observed for nominal $Ni_{33}Co_{17}Mn_{35}Ti_{15}$.

Zero-field DSC measurements have been performed to analyze the temperature dependence of the transition entropy change associated with the martensitic transformation in Ni(-Co)-Mn-Ti. The heat flow data is shown for $Ni_{35}Co_{15}Mn_{50-y}Ti_y$ and $Ni_{33}Co_{17}Mn_{50-y}Ti_y$ in figure 2 (b) and (e), respectively. The forward and reverse martensitic transformation are clearly visible upon cooling and heating as exo- and endothermic peaks, respectively. The transition entropy change is calculated based on the area below the heating curves (see equation 5). Qualitatively, one can clearly see an increase in $\Delta s_t$ with an increasing martensitic transformation temperature and consequently reduction of Ti content, $\Delta M$ and $T_C^A - T_t$ in both series.

The transition entropy change of the martensite to austenite transformations are shown for $Ni_{35}Co_{15}Mn_{50-y}Ti_y$ and $Ni_{33}Co_{17}Mn_{50-y}Ti_y$ in figure 2 (c) and (f), respectively. In both series, one can observe a very good agreement between the various methods used to determine $\Delta s_t$ as well as literature values [35, 37, 39, 42, 44]. The competition of positive lattice entropy change $\Delta s_{lat}$ and towards lower temperatures increasingly negative magnetic entropy change $\Delta s_{mag}$ leads to the overall decrease of $\Delta s_t = \Delta s_{lat} + \Delta s_{mag}$ for $T_t \leq T_C^A$. This competition leads to the compensation of both entropy change contributions at $T_{comp}$, which is reached at around 75 K and 300 K for the $Ni_{35}Co_{15}Mn_{50-y}Ti_y$ and $Ni_{33}Co_{17}Mn_{50-y}Ti_y$ series, respectively. These compensation temperatures are in excellent agreement with the isofield magnetization curves shown in figure 2 (a, d), as no transformations can be observed below the respective $T_{comp}$. The substantial difference in $T_{comp}$ between both series originates from the larger negative magnetic entropy change contribution in $Ni_{33}Co_{17}Mn_{50-y}Ti_y$ at any given temperature due to increased ferromagnetic coupling and saturation magnetization in the austenite phase caused by the increased Co content.

If $T_t \geq T_C^A$, $\Delta s_{mag}$ approaches zero due to the absence of ferromagnetic order in the austenite phase and the transition entropy change is purely defined by the structural subsystem contribution $\Delta s_{lat}$ [49], showing a value of approximately 65 J(kgK)$^{-1}$. Since $\Delta s_t$ is equivalent for both series at $T_t \geq T_C^A$, $\Delta s_{lat}$ is independent on such minor adjustments in chemical composition. In order to estimate $\Delta s_{lat}$ for an extended composition range in Ni(-Co)-Mn-Ti Heusler alloys, $Ni_{47.9}Mn_{34.2}Ti_{17.9}$ has been analyzed (see supplementary material S3), as the composition is substantially different to the $Ni_{35}Co_{15}Mn_{50-y}Ti_y$ and $Ni_{33}Co_{17}Mn_{50-y}Ti_y$ series and the absence of Co guarantees non-ferromagnetic order in martensite and austenite state [47]. The detected transition entropy change of 66.2 J(kgK)$^{-1}$ indicates the independence of $\Delta s_{lat}$ on the Co content and represents thereby the upper limit of $\Delta s_t$ in all-d-metal Ni(-Co)-Mn-Ti Heusler alloys. This value is also in good agreement with 66.1 J(kgK)$^{-1}$ reported for nominal $Ni_{50}Mn_{31.75}Ti_{18.25}$ in reference [40]. Based on the apparent independence of $\Delta s_{lat}$ on such large differences in Co and Ti content, the transition entropy change at temperatures above $T_C^A$ can be assumed to be also constant in the $Ni_{35}Co_{15}Mn_{50-y}Ti_y$ and $Ni_{33}Co_{17}Mn_{50-y}Ti_y$ series. However, the exact behavior remains unclear as samples with a lower Ti content than approximately 9 at.% form two distinctively different Ti-rich and Ti-poor phases with individual martensitic transformation temperatures according to their $e/a$ ratio (see supplementary material S4), causing the detectable heat flow of these transformations to be unreliable for $\Delta s_t$ determinations due to parasitic contributions of the respective non-transforming phase. Interestingly, the chemically homogeneous single phase samples with $T_t \geq T_C^A$ show similar transition entropy changes $\Delta s_t = \Delta s_{lat}$ of approximately 65 J(kgK)$^{-1}$, even though different martensite crystal structures are present. This is evident by comparing modulated orthorhombic 4O(IC) Ni$_{47.9}$Mn$_{34.2}$Ti$_{17.9}$ ($a = 4.37$ Å, $b = 5.50$ Å, $c = 4.27$ Å, $\beta = 90°$, $q = (0, 0, 0.498)$ at 300 K, see supplementary material S3) with almost identical non-modulated L1$_0$ of Ni$_{33.4}$Co$_{14.6}$Mn$_{42.8}$Ti$_{9.2}$ ($a = 3.64$ Å, $c = 7.27$ Å at 296 K, see figure 1 (a)) and Ni$_{31.6}$Co$_{16.4}$Mn$_{42.8}$Ti$_{9.2}$ ($a = 3.64$ Å, $c = 7.25$ Å at 296 K, see figure 1 (d)).

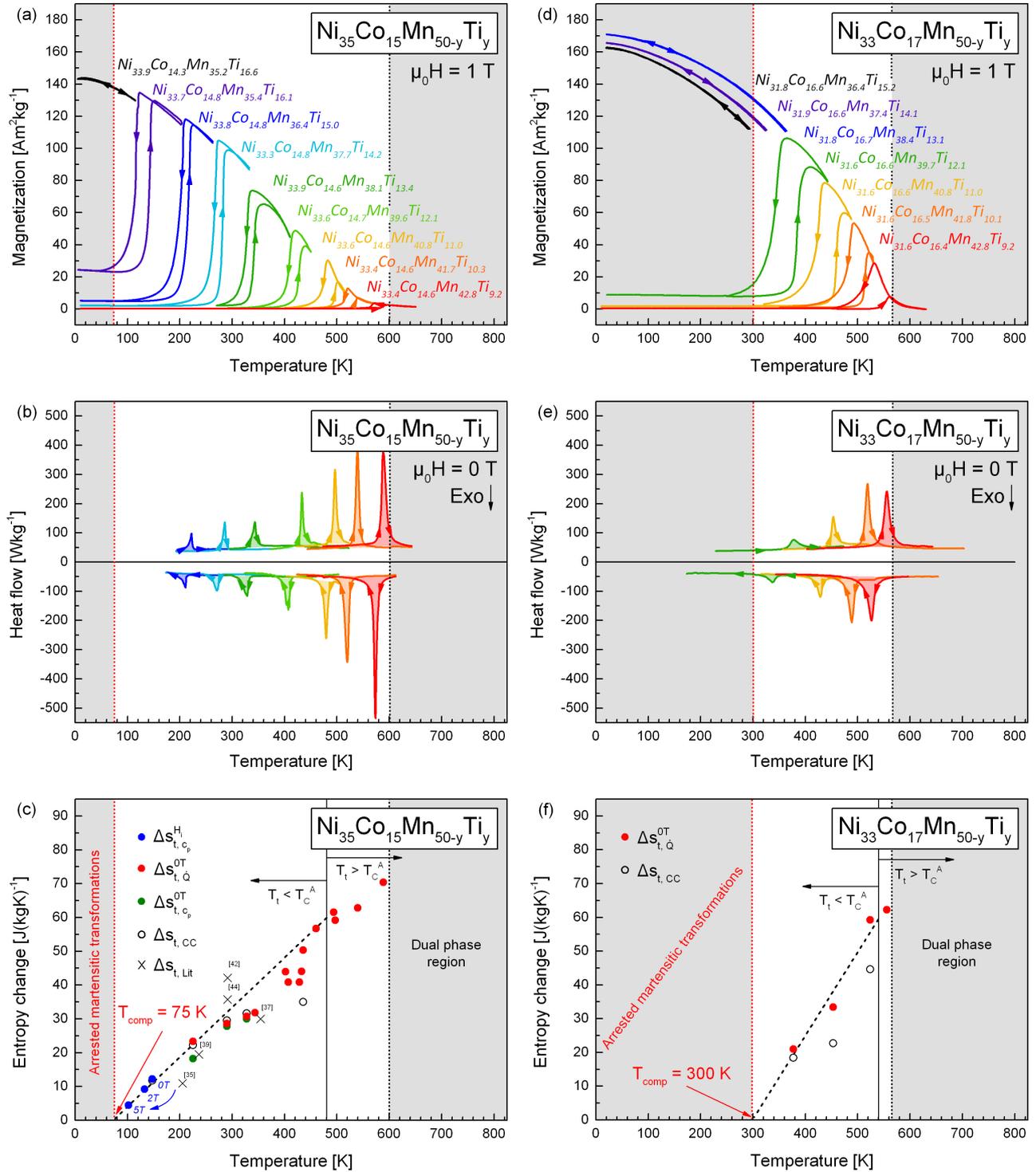

**Figure 2:** Isofield magnetization curves in 1 T (a, d), zero-field heat flow curves (b, e) and transition entropy changes (c, f) of the nominal $Ni_{35}Co_{15}Mn_{50-y}Ti_y$ (a, b, c) and $Ni_{33}Co_{17}Mn_{50-y}Ti_y$ (d, e, f) series. All samples are labeled with the composition determined by EDX. (b, e) Shaded areas under the heat flow curves qualitatively illustrate the transition entropy change. (c, f) The transition entropy change is determined by PPMS-14T $c_p$ measurements (blue), sapphire-calibrated DSC $c_p$ measurements (green), DSC heat flow measurements (red) and the Clausius-Clapeyron equation (open circles). Literature values [35, 37, 39, 42, 44] (crosses) are given for comparison. Dashed lines are drawn to guide the eye. The compensation temperatures $T_{comp}$ can be estimated as 75 K and 300 K for the $Ni_{35}Co_{15}Mn_{50-y}Ti_y$ and $Ni_{33}Co_{17}Mn_{50-y}Ti_y$ series, respectively. The estimated temperature at which the martensitic transformation temperature $T_t$ is equal to the austenite Curie temperature $T_C^A$ is given as vertical solid line. Shaded areas at low and high temperatures illustrate regions with arrested martensitic transformations and chemically inhomogeneous dual phase microstructures, respectively.

Compared to classical Heusler alloys with an inverse magnetocaloric effect [49–55], compensation temperatures of 75 K and 300 K represent one of the lowest and highest $T_{comp}$ values, underlining the versatility of the all-d-metal Ni(-Co)-Mn-Ti system. The compensation temperature of 300 K in Ni$_{33}$Co$_{17}$Mn$_{50-y}$Ti$_y$ also clearly indicates that arrested martensitic transformations are of thermodynamic and not kinetic origin. Transition entropy changes of 65 J(kgK)$^{-1}$ at $T_t \geq T_C^A$ are substantially larger compared to classical Heusler alloys with typical values of approximately 30 J(kgK)$^{-1}$ for Ni-Mn-Ga [77, 78], 45 J(kgK)$^{-1}$ for Ni-Mn-In [49] and Ni-Mn-Sn [79], 50 J(kgK)$^{-1}$ for Ni-Co-Mn-Sn [80] and 55 J(kgK)$^{-1}$ for Ni-Co-Mn-In [49].

In combination with the phase diagram of Ni-Co-Mn-Ti published in reference [38], the temperature dependence of the transition entropy change (see figure 2 (c, f)) allows the design of samples with tailored properties for any given caloric cooling application by tuning the Ni/Co and Mn/Ti ratio. For the following simplified illustration of the underlying universal design concept based on $\Delta s_t$ in the transition temperature range $T_{comp} \leq T_t \leq T_C^A$, a constant non-negligible transition width $A_f - A_s$ is assumed. At any fixed $T_t$, samples with a higher Co content show smaller $\Delta s_t$ and larger $\Delta M$ values, which result according to Clausius-Clapeyron (see equation 3) in an increased magnetic field sensitivity of the phase transition $dT_t/\mu_0 dH$. Therefore, such samples achieve large non-saturated magnetocaloric effects in moderate magnetic field changes up to 2 T due to the enhanced $dT_t/\mu_0 dH$. However, the saturated magnetocaloric effect will be limited due to the comparatively small $\Delta s_t$. Contrary, at the same $T_t$, samples with a small Co content show larger $\Delta s_t$ and smaller $\Delta M$ values, leading to larger saturated magnetocaloric effects in sufficiently large magnetic fields to overcome the small magnetic field sensitivity of the phase transformation. This combination of an enhanced saturated magnetocaloric effect and an improved mechanical stability [39, 40] can be ideally utilized in the multicaloric cooling cycle [16], giving the all-d-metal Ni-Co-Mn-Ti Heusler alloys an advantage over classical Heusler alloys, which suffer from smaller saturated magnetocaloric effects due to smaller $\Delta s_t$ values and an increased brittleness [81].

The decreasing transition entropy change with increasing magnetic field shown for Ni$_{33.7}$Co$_{14.8}$Mn$_{35.4}$Ti$_{16.1}$ (see figure 2 (c), solid blue circles) demonstrates the equivalence of chemical composition and magnetic fields with respect to the shift of $T_t$ towards and below $T_{comp}$, which will be discussed in greater detail in the context of figure 3 and 4.

### 3.3. Arrested martensitic transformations

To analyze the responses of the magnetic, structural and electronic subsystems to the temperature- and magnetic field-induced martensitic transformation close to and below the compensation temperature in Ni$_{35}$Co$_{15}$Mn$_{50-y}$Ti$_y$, simultaneous measurements of magnetization $M$, strain $\Delta L/L_0$ and electrical resistivity $\rho$ have been carried out for Ni$_{33.7}$Co$_{14.8}$Mn$_{35.4}$Ti$_{16.1}$. The isofield and isothermal curves are shown in figure 3 (a) and (b), respectively. One can see in both figures that martensite is the low-temperature, low-magnetization and high-resistivity phase. The strain, averaged over two strain gauges, is in good agreement with [47] as well as temperature-dependent XRD measurements (see supplementary material S5), which show an average isotropic strain for a non-textured polycrystalline sample of approximately 0.4% during the transition from modulated monoclinic 6M(IC) martensite ($a = 4.34$ Å, $b = 5.46$ Å, $c = 4.24$ Å, $\beta = 93.4°$, $q = (0, 0, 0.345)$ at 20 K) to cubic B2 austenite ($a = 2.950$ Å at 300 K). The small deviation with respect to the strain value arises from an optimized microstructure with large grain sizes due to abnormal grain growth during heat treatment [38]. The increased electrical resistivity in the martensite phase is due to the formation of twin boundaries (see figure 3 (b), insets), which are increasing the electron scattering cross-section. The absolute $\rho$ values are similar to nominal Ni$_{35}$Co$_{15}$Mn$_{35}$Ti$_{15}$ [47] and Ni$_{35}$Fe$_{15}$Mn$_{35}$Ti$_{15}$ [82].

The temperature dependence of all subsystems (see figure 3 (a)) shows the stabilization of the high-temperature ferromagnetic austenite by magnetic field, leading to the observable decrease of the transition temperature with increasing field strength. The amount of residual austenite at low temperatures grows with increasing magnetic field since the distribution of locally varying transition temperatures, caused by minor chemical inhomogeneities on the microscale, is increasingly shifted below $T_{comp}$ [49]. This is evident by the simultaneously measured high magnetization and strain value as well as the low electrical resistivity after field cooling. This trend continues until the martensitic transformation becomes fully arrested in fields larger than 5 T. The temperature at which the martensitic transformation is suppressed is in good agreement with the compensation temperature of 75 K, determined based on the compositional influence on $\Delta s_t$ in the Ni$_{50-x}$Co$_x$Mn$_{50-y}$Ti$_y$ series in figure 2 (c), demonstrating the equivalent influence of chemical composition and magnetic field in this context. The isothermal magnetization curves (see figure 3 (b)) show field-induced transformations down to 10 K since each measurement temperature is approached in zero-field. The ferromagnetic background in the magnetization signal at low magnetic fields indicates the presence of residual austenite. Isofield and isothermal measurements show a substantially increased thermal and magnetic hysteresis width at low temperatures, respectively.

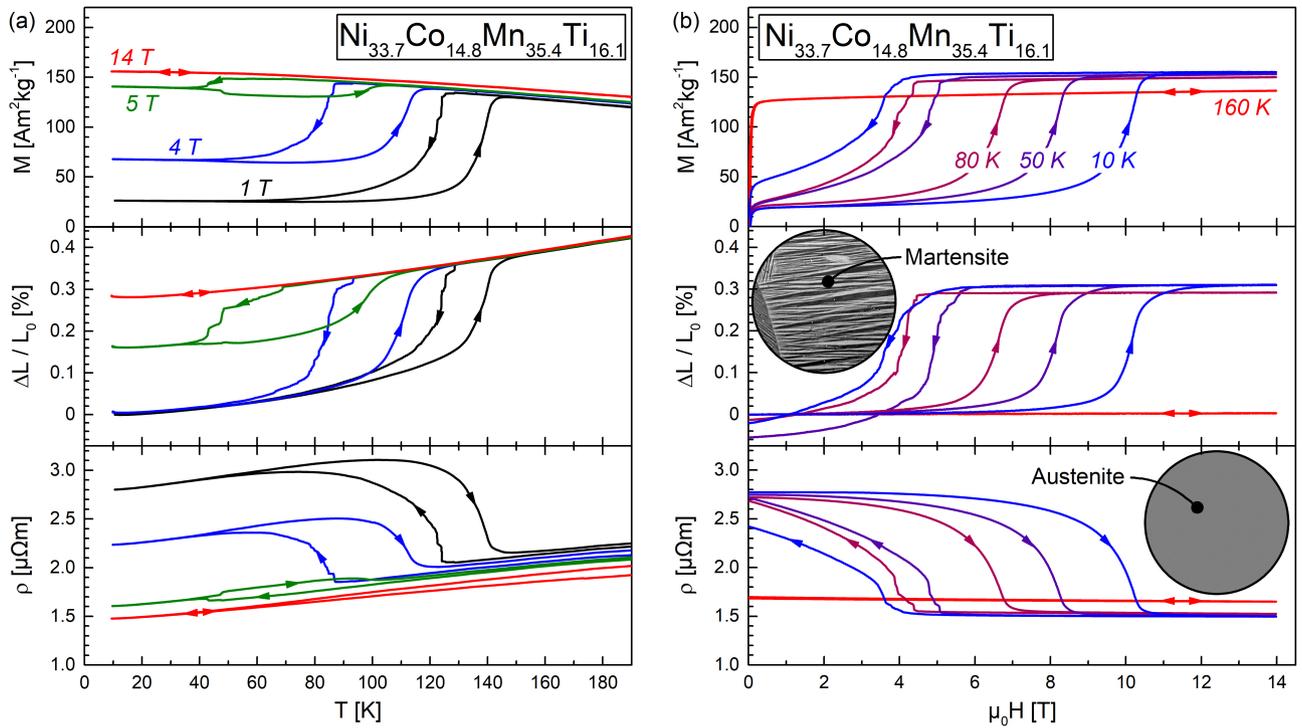

**Figure 3:** Isofield (a) and isothermal (b) simultaneous measurements of magnetization $M$, strain $\Delta L/L_0$ and electrical resistivity $\rho$ of the $Ni_{33.7}Co_{14.8}Mn_{35.4}Ti_{16.1}$ sample. (a) Temperature dependencies are shown for 1, 4, 5, and 14 T. (b) Magnetic field dependencies are shown for 10, 50, 80, and 160 K. Exemplary representations of the martensite and austenite microstructure are given with micrographs imaged by optical microscopy, positioned at low and high magnetic fields, respectively.

The results regarding the temperature- and magnetic field-dependent hysteresis in $Ni_{33.7}Co_{14.8}Mn_{35.4}Ti_{16.1}$ have been summarized in a phase diagram in figure 4 (a). The phase diagram is based on the martensite start $M_s$ and austenite finish $A_f$ points, determined via the double tangent method. Martensite finish $M_f$ and austenite start $A_s$ are omitted for clarity. The phase diagram shows an excellent agreement of isothermal and isofield measurements regarding the start and finish points of the transformation. The arrested martensitic transformation can clearly be recognized at the maximum of $M_s$, as this indicates the magnetic field beyond which the transformation cannot be induced upon cooling. The magnetic hysteresis width $H_{Hys}$ shows an abnormal behavior at $T \leq T_{comp}$, as the forward martensitic transformation is shifted towards smaller magnetic fields with decreasing temperature. This highly non-linear and non-monotonic shift is in contrast to the high-temperature region (90 K $\leq T \leq$ 150 K) in which both $A_f$ and $M_s$ shift linear towards higher magnetic fields with decreasing temperature. The broadening of hysteresis and the correlated increase of dissipation energy due to the abnormal behavior of the starting point of the field-induced transition from high-temperature to low-temperature phase upon field removal at $T \leq T_{comp}$ appears to be a general feature for a variety of magnetocaloric materials, such as classical Ni-Mn-based Heusler alloys [50–52] and Fe-Rh [34].

To analyze the field dependence of the transition entropy change $\Delta s_t$ close to $T_{comp}$, heat capacity measurements have been carried out for $Ni_{33.7}Co_{14.8}Mn_{35.4}Ti_{16.1}$ and the results are shown in figure 4 (b). The dataset is consistent with figure 3 as no transformations can be induced by temperature in magnetic fields above 5 T. The determined transition entropy changes in 0, 2, and 5 T (see figure 2 (c), solid blue circles) gradually decrease as less material transforms due to the shift of $T_t$ below $T_{comp}$. The $s(T)$ diagram shows an excellent agreement with $\Delta s_T$ determined with Maxwell's relation (see equation 2) and only small isothermal entropy changes $\Delta s_T$ and adiabatic temperature changes $\Delta T_{ad}$ can be observed as $T_t$ is close to the compensation temperature. Additionally, the heat capacity measurements clearly reveal the continuous and discontinuous character of the reverse and forward martensitic transformation upon heating and cooling, respectively [83].

Simultaneous isofield measurements of magnetization and strain for $Ni_{31.6}Co_{16.6}Mn_{39.7}Ti_{12.1}$ reveal the same behavior as in $Ni_{33.7}Co_{14.8}Mn_{35.4}Ti_{16.1}$ with respect to a growing amount of residual austenite after cooling with increasing magnetic fields, caused by the gradual shift of $T_t$ below $T_{comp}$ (see supplementary material S6). However, due to the smaller $dT_t/\mu_0 dH$ value of -5.1 KT$^{-1}$ compared to -9.1 KT$^{-1}$ of $Ni_{33.7}Co_{14.8}Mn_{35.4}Ti_{16.1}$, a field of 14 T is insufficient to completely suppress the phase transition and observe the broadening of hysteresis due to the abnormal behavior of $M_s$ at $T \leq T_{comp}$.

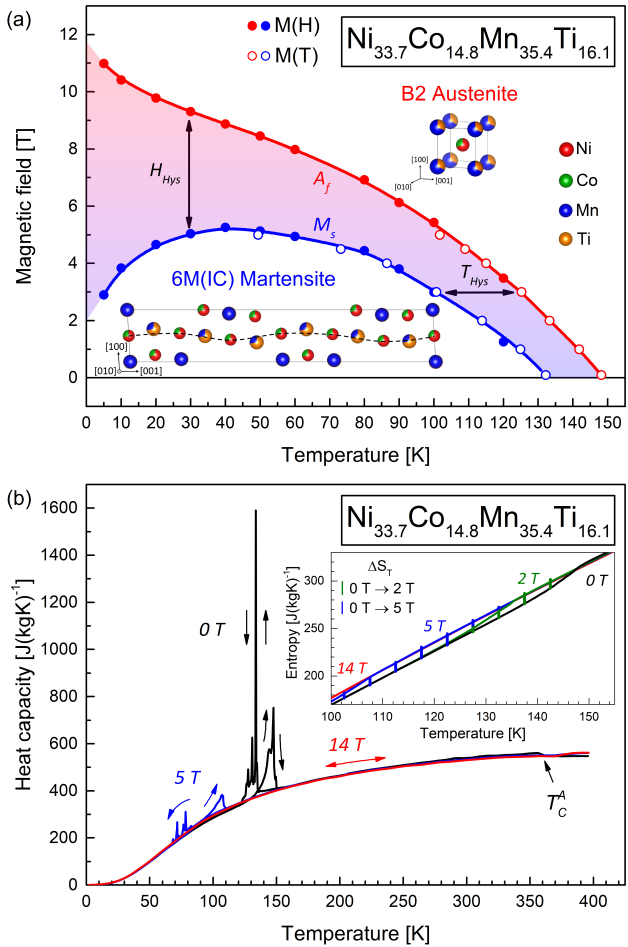

**Figure 4:** (a) Temperature- and magnetic field-dependent phase diagram of the $Ni_{33.7}Co_{14.8}Mn_{35.4}Ti_{16.1}$ sample with the corresponding crystal structures of the high-temperature B2 austenite and low-temperature 6M(IC) martensite phase, including an illustration of the incommensurate modulation of the atomic positions (dashed line). The austenite finish $A_f$ (red) and martensite start temperature $M_s$ (blue), defining the magnetic and thermal hysteresis width $H_{Hys}$ and $T_{Hys}$, are determined based on $M(H)$ (solid circles) and $M(T)$ (open circles) measurements, shown in figure 3 (b) and (a), respectively. (b) Temperature- and magnetic field-dependent heat capacity measurements of $Ni_{33.7}Co_{14.8}Mn_{35.4}Ti_{16.1}$. The inset shows the $s(T)$-diagram based on the $c_p$ measurements during heating. Vertical lines represent isothermal entropy changes calculated by Maxwell's relation (see equation 2).

### 3.4. Pulsed field measurements of $Ni_{33.7}Co_{14.8}Mn_{35.4}Ti_{16.1}$

To study the magnetocaloric effect as well as the impact of hysteresis on $\Delta T_{ad}$ at temperatures close to and below $T_{comp}$, adiabatic temperature change measurements have been carried out for $Ni_{33.7}Co_{14.8}Mn_{35.4}Ti_{16.1}$ in pulsed magnetic fields. The adiabatic temperature change upon field-induced martensite to austenite transformation is shown in figure 5 (a) for a variety of magnetic field pulses at temperatures of 15, 90, and 170 K.

At 170 K, a small positive $\Delta T_{ad}$ is observable due to the conventional second-order magnetocaloric effect of the ferromagnetic austenite, which is the only phase being present at this temperature (see figure 3 (a)). The pulsed field partly aligns the magnetic moments in adiabatic conditions, causing the magnetic entropy to decrease, which is counterbalanced by an increasing lattice entropy since the total entropy stays constant. This results in the apparent monotonic increase of $\Delta T_{ad}$ with increasing magnetic field strength.

At 90 K, the pulsed magnetic field induces the martensite to austenite transformation as the ferromagnetic austenite is stabilized by the magnetic field. This transformation is accompanied by a small negative $\Delta T_{ad}$, as the magnetocaloric effect is of inverse character and the initial temperature prior to the pulse is close to the compensation temperature of 75 K in the $Ni_{35}Co_{15}Mn_{50-y}Ti_y$ series (see figure 2 (c)). The inverse magnetocaloric effect, associated with the field-induced transformation, saturates at approximately -2.8 K in 20 T as the transformation is completed (see figure 3 (b) and figure 5 (a)). At 90 K and 170 K, adiabatic temperature changes measured in pulsed fields are consistent with indirectly determined $\Delta T_{ad}$ values based on $c_p$, demonstrating good comparability of both measurement techniques at such elevated temperatures.

At 15 K, however, a seven times larger and positive $\Delta T_{ad}$ of 15 K in 20 T upon completely induced martensite to austenite transformation can be observed compared to 90 K. Additionally, the directly measured $\Delta T_{ad}$ is strongly contradictory to the expected small negative $\Delta T_{ad}$ determined based on $c_p$ measurements shown in figure 4 (b). However, the adiabatic temperature change measured in pulsed fields is strongly coupled to the induced magnetostructural phase transformation and not due to artificial signal contributions, which is apparent by comparing the simultaneously measured thermal and magnetic response of the sample as a function of magnetic field (see figure 5 (a), inset). This clearly indicates the presence of an additional effect dominating the thermal response of the material at such low temperatures. The origin of the change in sign and irreversibility of $\Delta T_{ad}$ is an increased dissipation energy $q_{diss} = \mu_0 \oint H dM$ (see supplementary material S7) during the phase transformation at such low temperatures due to an increased magnetic hysteresis width (see figure 4 (a)), leading in combination with an extremely small specific heat capacity (see figure 4 (b)) to a significant positive thermal response of the material $\Delta T_{diss} = 0.5 \cdot q_{diss} c_p^{-1}$. The factor 0.5 considers that dissipation losses occur both during field application and removal. To include the influence of adiabatic conditions during the magnetic field pulse on $\Delta T_{diss}$, an upper and lower limit of $\Delta T_{diss}$ is calculated based on the $q_{diss}$ and $c_p$ values at the start and finish conditions of the field-induced martensite to austenite transformation, respectively. The overall adiabatic temperature change $\Delta T_{ad}$ is thereby the sum of a small negative inverse magnetocaloric effect $\Delta T_{s(T)}$, which can be estimated with the $s(T)$ diagram based on $c_p$ measurements, and a large positive dissipation loss $\Delta T_{diss}$.

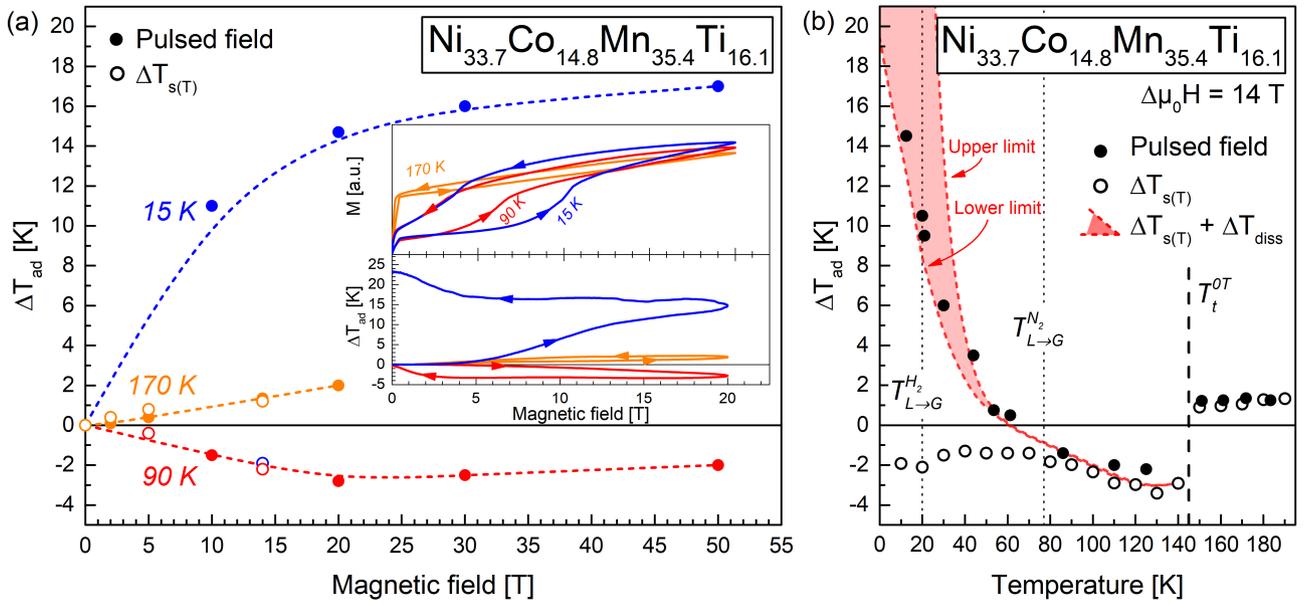

**Figure 5:** (a) Adiabatic temperature change $\Delta T_{ad}$ in $Ni_{33.7}Co_{14.8}Mn_{35.4}Ti_{16.1}$ at 15, 90, and 170 K. The $\Delta T_{ad}$ values are based on direct measurements in pulsed magnetic fields (solid circles) and indirect determinations with $s(T)$ based on $c_p$ measurements ($\Delta T_{s(T)}$, open circles). For clarity, $\Delta T_{ad}$ values are only given for the maximum magnetic field strength of each pulse. Dashed lines are drawn to guide the eye. The inset shows the magnetic field dependence of the simultaneously measured magnetization (top panel) and adiabatic temperature change (bottom panel) in a pulsed magnetic field of 20 T at 15, 90, and 170 K. (b) Temperature dependence of the adiabatic temperature change measured with pulsed magnetic fields (solid circles) and determined with $s(T)$ based on $c_p$ measurements ($\Delta T_{s(T)}$, open circles) in a magnetic field change of 14 T. The estimated effect of dissipation losses $\Delta T_{diss}$ on the resulting adiabatic temperature change is given by $\Delta T_{ad} = \Delta T_{s(T)} + \Delta T_{diss}$. The upper and lower limit (red dashed lines) of the estimation is calculated with $q_{diss}$ and $c_p$ values at the start and finish conditions of the adiabatic field-induced martensite to austenite transformation. Boiling temperatures of $H_2$ and $N_2$ are given by $T_{L \to G}^{H_2}$ and $T_{L \to G}^{N_2}$, respectively (dotted vertical lines). The martensite to austenite transformation temperature in 0 T is given by $T_t^{0T}$ (dashed vertical line).

Figure 5 (b) demonstrates that the measured adiabatic temperature change of $Ni_{33.7}Co_{14.8}Mn_{35.4}Ti_{16.1}$ can be very well described by $\Delta T_{ad} = \Delta T_{s(T)} + \Delta T_{diss}$. At low temperatures, $\Delta T_{diss}$ rises drastically, causing the estimation of the temperature change solely based on indirect methods using datasets obtained under isothermal or isofield conditions to break down as $\Delta T_{diss}$ dramatically influences $\Delta T_{ad}$. This is especially critical as it is common practice in low-temperature magnetocalorics to estimate $\Delta T_{ad}$ either based on the $s(T)$ diagram [84] or on $\Delta s_T$ and $c_p$ [85, 86]. Therefore, it is imperative to directly measure and consequently verify $\Delta T_{ad}$ for materials showing field-induced first-order phase transitions with non-negligible hysteresis at such low temperatures in the future. However, it should be noted that the precise determination of $\Delta T_{diss}$ and $\Delta T_{ad}$ is challenging at $T \leq 35$ K as only minor deviations in $c_p$ dramatically influence $\Delta T_{diss}$ (see figure 5 (b)). According to the formula used to estimate $\Delta T_{diss}$, the assumption of a constant $c_p$ itself ultimately leads to a divergent behavior of the upper limit of $\Delta T_{diss}$ at $T \to 0$, since $\Delta T_{diss} \to \infty$ if $c_p \to 0$. Nevertheless, taking into account the conditions prior to as well as subsequently to the adiabatic magnetic field-induced martensite to austenite transformation, it is possible to give a reasonable estimation of $\Delta T_{ad}$ (see figure 5 (b)).

The origin of the positive $\Delta T_{ad}$ is therefore not linked to a change from inverse to conventional magnetocaloric effect, which could be proposed based on a linear extrapolation of $\Delta s_t(T)$ to $T < T_{comp}$ in figure 2 (c), resulting in a reversed sign of $\Delta s_t$ and consequently $\Delta T_{ad}$. This scenario is unlikely as the adiabatic temperature change upon field removal, i.e. the austenite to martensite transformation, is positive and has not changed sign (see figure 5 (a), inset). Furthermore, the same argument should hold true not only for the magneto- but also elastocaloric effect and not only in the $Ni_{35}Co_{15}Mn_{50-y}Ti_y$ but also in the $Ni_{33}Co_{17}Mn_{50-y}Ti_y$ series. However, the uniaxial stress-induced austenite to martensite transformation in $Ni_{31.9}Co_{16.6}Mn_{37.4}Ti_{14.1}$, which shows an arrested transformation below $T_{comp}$ of 300 K (see figure 2 (d)), clearly exhibits an as-expected conventional elastocaloric effect under adiabatic conditions at 220.9 K (see supplementary material S8).

The change of sign and irreversibility of $\Delta T_{ad}$ at low temperatures is not only interesting for the fundamental understanding of the intertwined intrinsic and extrinsic parameters determining the thermal response of caloric materials to an applied stimulus, but also raise an important question with respect to recently emerging research interest in the utilization of magnetocaloric materials for gas liquefaction at low temperatures [56, 58]. Currently, one of the most prominent materials for such an application are $RT_2$-based Laves phases, with T and R being transition metal (T) and critical rare-earth (R) elements, respectively. Intermetallics such as $RCo_2$ [84, 85] and $RAl_2$ [87, 88] show highly tuneable phase transitions down to temperatures of only 10 K. Equivalent to room-temperature magnetocalorics, the utilization of giant magnetocaloric effects coupled to first-order magnetic or magnetostructural phase transitions would enhance performance and thereby overall feasibility of the technology, assuming the availability of sufficiently large magnetic field changes required to overcome the non-negligible hysteresis. However, the here reported and so far overlooked aspect of hysteresis for low-temperature magnetocalorics, leading to an irreversible and substantial absolute increase of $\Delta T_{ad}$ due to dissipation losses, significantly limits such an utilization, as $\Delta T_{diss}$ clearly dominates the thermal response at $T \leq 50$ K (see figure 5 (b)). In combination with similar irreversible, positive and directly measured $\Delta T_{ad}$ signals in inverse magnetocaloric Fe-Rh [34] at elevated temperatures, inverse elastocaloric Co-Cr-Al-Si [89] and conventional elastocaloric Ti-Ni-Cu-Al [90] as well as Ti-Ni-based, Cu-based and Ni-Mn-based shape memory alloys [91], we expect the effect to be neither material nor stimulus specific but fundamentally linked to phase transitions with hysteresis, underlining the universal importance of mastering hysteresis [92] in all caloric materials showing first-order phase transitions, especially at such low temperatures.

## 4. Conclusion

We present a systematic study on the transition entropy change $\Delta s_t$ in multicaloric all-d-metal Ni(-Co)-Mn-Ti Heusler alloys. The obtained composition dependence of $\Delta s_t$ is fundamental to the design of materials with tailored phase transition properties for any given future caloric cooling application. The transition entropy change associated with the structural subsystem $\Delta s_{lat}$ of 65 J(kgK)$^{-1}$ has been isolated by tuning the Ni/Co and Mn/Ti ratio and thereby modifying the martensitic transition temperature $T_t$ and austenite Curie temperature $T_C^A$, so that $T_t \geq T_C^A$. Since the here reported $\Delta s_t$ values are substantially larger compared to classical Heusler alloys, all-d-metal Ni(-Co)-Mn-Ti will perform superior in caloric cooling applications harnessing the full potential of the phase transition due to enhanced saturated caloric effects. For $T_t \leq T_C^A$, we have identified the competition of positive lattice and negative magnetic entropy change contributions, leading to lower $\Delta s_t$ with decreasing temperature. This intrinsic competition leads to compensation temperatures $T_{comp}$ of 75 K and 300 K, below which the martensitic transformations are arrested in $Ni_{35}Co_{15}Mn_{50-y}Ti_y$ and $Ni_{33}Co_{17}Mn_{50-y}Ti_y$, respectively.

Based on the composition dependence of $\Delta s_t$, we simultaneously measured the responses of the magnetic, structural and electronic subsystems close to and below the compensation temperature $T_{comp}$ in $Ni_{33.7}Co_{14.8}Mn_{35.4}Ti_{16.1}$ and $Ni_{31.6}Co_{16.6}Mn_{39.7}Ti_{12.1}$. The temperature dependence of all subsystems shows the stabilization by magnetic field of the high-temperature ferromagnetic austenite, leading to the gradual shift of $T_t$ below $T_{comp}$ until the martensitic transformation becomes fully arrested in sufficiently large fields. Isofield measurements show a substantially increased hysteresis and consequently dissipation energy at cryogenic temperatures, primarily due to the shift of the forward martensitic transformation towards smaller magnetic fields with decreasing temperature.

Simultaneous magnetization and adiabatic temperature change measurements of $Ni_{33.7}Co_{14.8}Mn_{35.4}Ti_{16.1}$ in pulsed magnetic fields reveal a change in sign of $\Delta T_{ad}$ with substantial positive and irreversible values up to 15 K at 15 K due to the combination of increased dissipation energy and decreased heat capacity. Therefore, the here reported irreversibility of $\Delta T_{ad}$ demonstrates a so far overlooked limitation of utilizing phase transitions with non-negligible hysteresis at cryogenic temperatures for magnetocaloric gas liquefaction. Furthermore, as dissipation losses are fundamentally linked to first-order phase transitions with hysteresis, the observed effect is of universal importance for all caloric cooling applications of first-order materials at cryogenic temperatures.

## Acknowledgements

We acknowledge financial support by the Deutsche Forschungsgemeinschaft (DFG) within the CRC/TRR 270 (Project-ID 405553726) and BEsT (Project-ID 456263705) as well as by the European Research Council (ERC) under the European Union's Horizon 2020 research and innovation programme (Grant No. 743116). We acknowledge the support of the HLD at HZDR, member of the European Magnetic Field Laboratory (EMFL).